# ON THE FEASIBILITY TO STUDY INVERSE PROXIMITY EFFECT IN A SINGLE S/F BILAYER BY POLARIZED NEUTRON REFLECTOMETRY

Revised 6/11/2013 17:01:00

Yu.N. Khaydukov[1,2], B. Nagy[3], J.-H. Kim[2], T. Keller[2], A. Rühm[4], Yu.V. Nikitenko[5], K.N. Zhernenkov[6], J. Stahn[7], L. Kiss[3], A. Csik[8], L. Bottyán[3], V.L. Aksenov[1,5,9]

[1] *Skobeltsyn Institute of Nuclear Physics, Moscow State University, Moscow, Russia*
[2] *Max-Planck Institute for Solid State Research, Stuttgart, Germany*
[3] *Wigner Research Centre for Physics, Hungarian Academy of Sciences, Budapest, Hungary*
[4] *Max-Planck-Institut für Intelligente Systeme (formerly Max-Planck-Institut für Metallforschung), Stuttgart, Germany*
[5] *Joint Institute for Nuclear Research, Dubna, Russia*
[6] *Ruhr-Universität Bochum, Bochum, Germany*
[7] *Paul Scherrer Institut, Villigen, Switzerland*
[8] *Institute for Nuclear Research, Hungarian Academy of Sciences, Debrecen, Hungary*
[9] *Petersburg Nuclear Physics Institute, Gatchina, Russia*

y.khaydukov@fkf.mpg.de

Here we report on a feasibility study aiming to explore the potential of Polarized Neutron Reflectometry (PNR) for detecting the inverse proximity effect in a single superconducting/ferromagnetic bilayer. Experiments, conducted on the V(40nm)/Fe(1nm) S/F bilayer, have shown that experimental spin asymmetry measured at $T = 0.5 T_C$ is shifted towards higher Q values compared to the curve measured at $T = 1.5 T_C$. Such a shift can be described by the appearance in superconducting vanadium of magnetic sub-layer with thickness of 7 nm and magnetization of +0.8 kG.

A proximity of a superconductor (S) and a ferromagnet (F) leads to the appearance of a great number of intriguing phenomena, such as spatial oscillation of electron density of states, π type S/F/S Josephson junctions, F/S/F spin valves, etc (see reviews 1-3). The study of these phenomena is of interest for our understanding of the physics of strongly correlated electron systems, and also of particular interest for creating a new generation of spintronic devices. One such phenomenon is the inverse proximity effect - the appearance of magnetic correlations in the superconductor (S) close to the interface in contact with the ferromagnet (F) [4,5]. The origin of this effect is the correlation of conduction electrons of the superconductor with free electrons in the ferromagnet, which, in turn, are exchange coupled. This correlation leads to the development of a finite magnetic moment in the superconductor. The depth distribution of the induced magnetic moment can be approximately written as $\delta M(z) \approx \pm \delta M(0) \cdot \exp(-z/\xi_S)$, where $\delta M(0)$ is the magnitude of induced moment on the S/F interface and $\xi_S$ is the superconducting coherence length. The sign of the induced magnetization can be either parallel or antiparallel to the direction of the vector of magnetization of the F layer, depending on the quality and the transparency of the interface, the thickness and the exchange field of the F layer [6,7], the magnetic state of the F layer [8,9], etc.

For the experimental investigation of the inverse proximity effects different magnetometric methods have been used. SQUID magnetometry was used [10] to study the magnetic state of a [Nb(100nm)/Py(10nm)]$_2$ system. An increase of the magnetic moment was observed at $T < T_C$. The maximum effect was observed at the experimentally reached minimum temperature. In zero field and in $H = 100$ Oe $\Delta m/m_0$ was found to be 6% and 5%, respectively (here $\Delta m \equiv m(T<T_c) - m_0$ - change of the total magnetic moment of the sample, $m_0$ – magnetic moment of the F layer above $T_c$). The authors explain the increase of the magnetic moment by a Meissner response of the S layer on the stray field produced by F layer. Nuclear Magnetic Resonance (NMR) was used in articles [11] and [12] to detect inverse proximity effect in Ni/V(30-70nm)/Ni/MgO systems. However, NMR inherently needs the application of DC and AC magnetic fields which makes comparison of the experiment with the theory rather ambiguous. The authors give a rough estimate $\delta M(0) \sim 3$kGs of the induced magnetization of the S/F interface. In the same issue of Physical Review Letters, another article [13] reported upon the observation of the inverse proximity effect in Pb/Ni and Al/CoPd bilayers using the polar magnetooptical Kerr effect. As a requirement of the technique, the authors had to apply a special procedure to magnetize the F layer perpendicular to the sample surface. Below $T_C$ they observed a finite magnetic signal indicative of a magnetic moment opposite to the F magnetization. No estimates of the value of induced magnetization are given in the work. The above mentioned experimental methods are highly sensitive, but lacking the required depth selectivity. Consequently, the thickness of the induced magnetic sublayer cannot be explored and compared with the theoretical predictions. An alternative experimental method to study the changes in the magnetization state of a sublayer is polarized neutron reflectometry (PNR). PNR measures the nuclear and the magnetic depth-profiles, the latter via the interaction of neutron spins with the magnetic induction in the system. Advantages of the neutrons for such studies are the very low absorption in the layers of the required thickness for superconductivity which allows studying deeply buried S/F interfaces. Another advantage of PNR is the possibility to measure in a low-intensity (several Oersteds) and DC-only magnetic fields with arbitrary to the sample plane direction. This allows to avoid problems with data misinterpretation due to the Meissner effect, vortex state etc. To increase magnetic PNR signal, usually multilayered systems are used where the investigated S/F bilayer is repeated several times [14,15]. Deposition of multilayered S/F structures may inevitably lead to an increased cumulative roughness of the S/F interface. Since the magnitude and even the sign of the induced magnetization strongly depends on the quality of S/F interface, this may lead to the situation when induced magnetization is different for the different S/F interfaces. This greatly complicates the interpretation of the PNR data. In order to avoid this complication we suggest studying the inverse proximity effect on the simplest S/F system – a single S/F bilayer. To enlarge small magnetic scattering in this case waveguide enhancement of the neutron standing waves [16,17] can be used. The advantage of the method is the strong enhance of the intensity of the magnetic scattering (gain factor is of order $10^1$-$10^2$). The main disadvantage is the small depth selectivity.

Waveguide enhancement was already used to study magnetic proximity effects in Cu(40nm)/V(40nm)/Fe(1nm)/MgO system with V/Fe bilayer of S/F type [18]. A 40% increase of the total magnetic moment of the S/F bilayer was observed below $T_C$. Using only the waveguide regime we were not able to determine whether the observed growth of magnetic moment is associated with the appearance of the induced magnetization in the S layer or with the increase of the magnetization in the F layer. In this article we will discuss how such information can be derived from the comparison of the neutron spin asymmetries above and below $T_C$.

The sample of Cu(32nm)/V(40nm)/Fe(1nm)/MgO(001) nominal composition was prepared using molecular beam epitaxy in the Wigner Research Centre for Physics. Here V(40nm) is a conventional BCS superconductor and Fe(1nm) is conventional itinerant ferromagnet. The high quality of the layers and S/F interface was proved by several techniques including Secondary Neutral Mass Spectrometry (SNMS, Fig. 1a) and (synchrotron) x-ray reflectometry [19]. In particular, the root-mean square roughness of the Fe/V interface was found to be less than 0.6 nm. The V/Cu interface, in contrary, is not smooth. The high intermixing of vanadium and copper leads to the creation of approximately 5 nm transition region.

Superconducting parameters of the S layer were defined using standard four-point DC measurements of the electrical resistivity with magnetic field applied parallel to the sample surface. Series of temperature and magnetic field scans allowed to define the critical temperature $T_C$ and upper critical field $H_{C2}$. Temperature dependence of the upper critical field $H_{C2}(T)$ is shown on the left inset to the Fig. 1b. It shows typical 2D superconductor behavior $(1-T/T_C)^{1/2}$ [20]. A fit of the experimental dependence $H_{c2}(T)$ allowed to define values $T_C = 3.4 \pm 0.1$ K and $\xi_S = 9.3 \pm 0.1$ nm. Another important superconducting parameter, the magnetic field penetration length $\lambda$, can be estimated from expression $H_{C2}(0) \approx 5\, H_{bulk}\, \lambda(0)/d_S$, where $H_{bulk} = 0.14$ T is the critical field for bulk vanadium, $d_S = 35$ nm – thickness of the S layer found from analysis of X-ray, neutron and SNMS data. The obtained value $\lambda(0) \approx 100$nm exceeds several times the thickness of the S layer. Hence, magnetic induction induced in S layer due to Meissner effect will not exceed a value of 1% of the applied magnetic field. Magnetic properties of the F layer were defined using SQUID magnetometry. The saturation magnetization and the coercivity at $T = 10$ K are $M_{sat} = 17.5$ kGs and $H_{coer} = 35$ Oe, respectively (see right inset to the Fig. 1b). The SQUID magnetometer was also used to define the temperature dependence of the magnetic moment in the vicinity of the superconducting phase transition (Fig. 1b). Before the cooling the F layer was saturated in a magnetic field $H = 1$ kGs and then cooled down below $T_C$ in a magnetic field $H = 10$ Oe. The measurement has shown that below $T_C$ a 40% increase of the magnetic moment takes place, which is consistent with the waveguide enhanced PNR data.

In order to define the reason of the increased magnetic moment we have measured the polarized neutron reflectivities above and below $T_C$ in conditions similar to SQUID. Measurements have been done on the angle dispersive reflectometers ADAM (ILL, France), NREX (FRM II, Germany) and time-of-flight reflectometer AMOR (PSI, Switzerland). Magnetic field of $H = 20$ Oe (ADAM, NREX) and $H = 100$ Oe (AMOR) was applied in-plane of the structure.

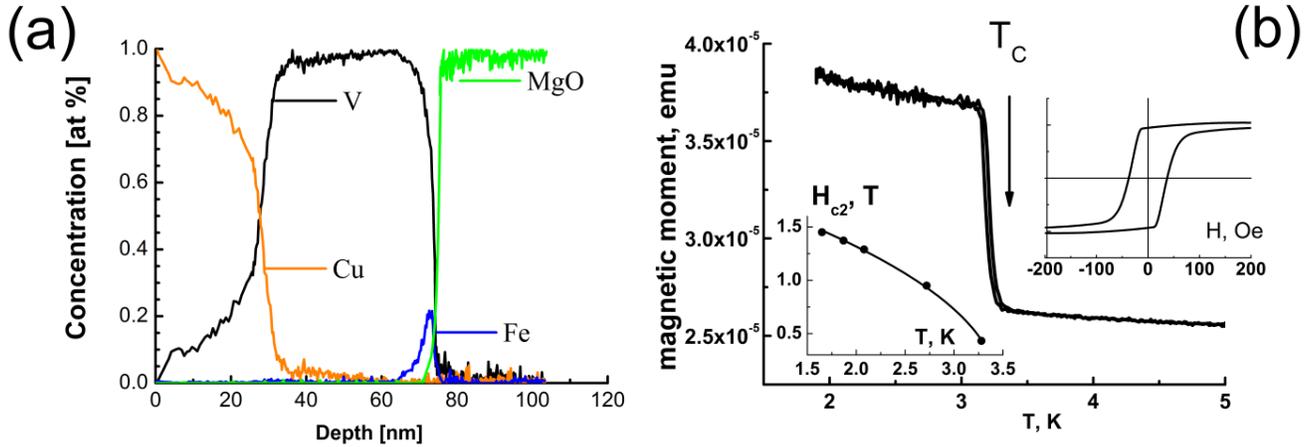

Fig. 1. (a) Depth profile of the concentrations of V, Cu, Fe and MgO measured by SNMS. (b) Temperature dependence of the magnetic moment around $T_C$, measured in magnetic field H = 10 Oe applied parallel to the sample surface. Left inset: temperature dependence of the upper critical field of the S layer (dots - experimental points, solid line - fit). Right inset: Magnetic hysteresis loop measured at T = 10 K.

The spin-up and spin-down specular neutron reflectivities $R^+(Q)$ and $R^-(Q)$ as a function of the momentum transfer Q were measured using position sensitive detectors (PSD). Since the magnetic contribution to the neutron scattering is small, it is useful to single it out using the so-called spin asymmetry $S(Q) \equiv [R^+(Q) - R^-(Q)]/[R^+(Q) + R^-(Q)]$. The spin asymmetry measured at T = 5K > Tc is depicted in Fig. 2a with the black line. This function was fitted to the model curve by a simple model, where only the F layer has magnetic signal. The best fit was obtained for the magnetization of the F layer of 14 kGs [18] which is consistent with the SQUID data. After cooling the sample below $T_C$ the shift of spin asymmetry towards higher Q-values was observed near Q = 0.35 nm$^{-1}$ (Fig. 2a). By controlling the position of the reflected neutron beam on PSD we confirmed that this shift is the intrinsic property of the sample but not an artificial fact caused by e.g. the lop-sided thermal expansion of the sample holder, shake of the cryostat during refilling etc. To describe the shift of the spin asymmetry below $T_C$ we have considered several models. The first model, change of the magnetic moment only in F layer cannot reproduce the shift of the spin asymmetry. In the second model we assumed the appearance of the magnetization in the vanadium layer below $T_C$ with exponential law: $\delta M(z) = \delta M(0) \exp[(z-z_1)]/\xi)$. Here $\delta M(0)$ is the amplitude of the induced magnetization on the S/F interface with coordinate $z_1$, $\xi$ – characteristic thickness of the magnetic sublayer. Parameters $\delta M(0)$ and $\xi$ were varied during the fitting. Position of the S/F interface $z_1 = 71$ nm was taken from the fit of the PNR data [18] and was fixed. Fit provides best parameters $\delta M(0) = 0.8\pm0.3$ kGs and $\xi = 7\pm1.0$ nm (Fig. 1b). To prove the sensitivity of the fit to the thickness of the sublayer, the dependence of the goodness-of-fit $\chi^2$ on $\xi$ is displayed in the inset of Fig. 2b. The resulting magnetic profile reproduces the 40% increase in magnetic moment detected with the help of waveguide enhanced PNR and SQUID measurements. This increased magnetization can be related to the inverse proximity effect, since the fitted thickness of the developed magnetic sublayer compared with the superconducting correlation length $\xi_S$ measured independently on the same sample. However, on this stage, we do not exclude that the induced magnetization may be related to the orbital effects, like, for example, pinning of the vortices on the S/F interface. This is indicative by the fact that the core of the vortex also positively magnetized and has dimension of the order of $\xi_S$. The magnetic field of the vortex core can be calculated as $B_c \approx \Phi_0 \ln(\lambda/\xi_S)/2\pi\lambda^2$, where $\Phi_0$ is magnetic flux quantum [21]. For our sample $B_c \approx 0.7$kGs, which is close to the value $\delta M(0)$ obtained from the fit of spin asymmetry. Detailed investigation will be presented elsewhere.

In conclusion, the feasibility of using the spin asymmetry of the neutron reflection for detecting the inverse proximity effect in a single superconducting/ferromagnetic bilayer is considered. The appearance in the superconductor of a new magnetic sub-layer will lead to a shift in the oscillation of the neutron spin asymmetry. First experiments have been conducted on a V(40 nm)/Fe(1 nm) S/F bilayer. Experiment has shown that the experimental spin asymmetry measured at T = 0.5 $T_C$ is shifted towards higher Q values compared to the curve at T = 1.5 $T_C$. Such shift can be described by appearance of a magnetic sub-layer in the superconducting vanadium with thickness of 7 nm and the magnetization of +0.8 kG. Since the thickness of the induced sub-layer is comparable with the coherence length of the S layer, appearance of this sub-layer can be attributed to the inverse proximity effect.

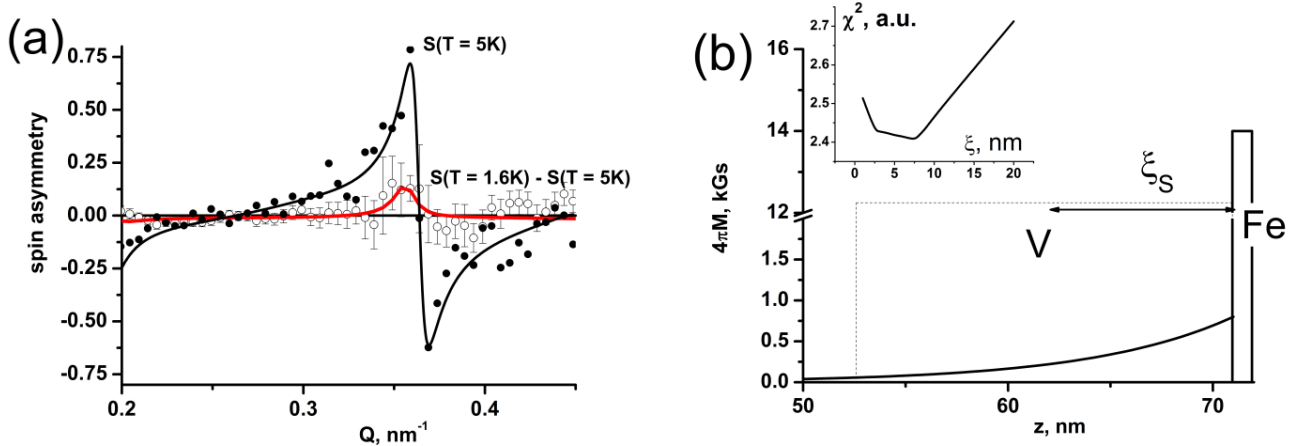

Fig. 2. (a) Experimental (dots) and model (solid line) difference of spin asymmetries $\Delta S \equiv S(0.5T_C) - S(1.5T_C)$. (b) Magnetic profile of the S/F bilayer below $T_C$ corresponding to the best-fit parameters $\delta M(0) = 0.8$ kGs and $\xi = 7$ nm. Horizontal arrow show the size of the superconducting correlation length found from transport measurements.

The authors would like to thank R. Salikhov, A. Vasenko and B. Keimer for the fruitful discussion of the results; H. Kolb and M. Beywl (FRM II TUM) for the assistance with the cryogenics during experiment at NREX reflectometer; A. Geresdi, S. Nieuwenhoven and the staff of the Low Temperature Solid State Physics Laboratory of the BUTE for their help in the AC resistance measurements; F. Tanczikó, G. Gy. Kertész and M. M. Dolgos for their work with sample preparation. B. N. would like to thank the financial support of the Forschungs-Neutronenquelle Heinz Maier-Leibnitz, Technische Universität München. The work was partially supported by RFBR grant No 12-02-12057 and TAMOP 4.2.2.A-11/1/KONV-2012-0036 project, which is co-financed by the European Union and European Social Fund.

## REFERENCES

1. Yu. A. Izyumov, Yu. N. Proshin and M. Khusainov, Phys. Usp. 45, 109 (2002)
2. K.B. Efetov, I.A. Garifullin, A.F. Volkov and K. Westerholt, Magnetic Heterostructures. Advances and Perspectives in Spinstructures and Spintransport, edited by H. Zabel and S.D. Bader, Series Springer Tracts in Modern Physics Vol. 227 (2007), p. 252.
3. A.I. Buzdin, Rev. Mod. Phys. 77, 935 (2005)
4. V. N. Krivoruchko and E. A. Koshina, Phys. Rev. B. 66, 014521 (2002)
5. F. S Bergeret, A. F. Volkov and K. B. Efetov, Phys. Rev. B. 69, 174504 (2004)
6. F. S. Bergeret, A. Levy Yeyati, and A. Martín-Rodero, Phys. Rev. B. 72, 064524 (2005).
7. M. Yu. Kharitonov, A.F. Volkov and K.B. Efetov, Phys. Rev. B 73, 054511 (2006)
8. J. Linder, T. Yokoyama and A. Sudbø, Phys. Rev. B 79, 054523 (2009)
9. N.G. Pugach, A.I. Buzdin, Appl. Phys. Lett. 101, 242602 (2012)
10. Hong-ye Wu, Jing Ni, Jian-wang Cai, Zhao-hua Cheng and Young Sun, Phys. Rev. B 76, 024416 (2007)
11. R.I. Salikhov, I. A. Garifullin, N.N. Garif'yanov et al, Phys. Rev. Let. 102, 087003 (2009)
12. R.I. Salikhov, I. A. Garifullin, N.N. Garif'yanov et al, Phys. Rev. B. 80, 214523 (2009)
13. Jing Xia, V. Shelukhin, M. Karpovski, A. Kapitulnik and A. Palevski, Phys. Rev. Let. 102, 087004 (2009)
14. J. Stahn, J. Chakhalian, Ch. Niedermayer, J. Hoppler et al., Phys. Rev. B. 71, 140509 (R) (2005)
15. D. K. Satapathy, M. A. Uribe-Laverde, I. Marozau et al, Phys. Rev. Lett. 108, 197201 (2012)
16. Yu.N. Khaydukov, Yu.V. Nikitenko, L. Bottyan, et al, Cryst. Rep. 55 (7),1235-1247 (2010)
17. Yu. N. Khaidukov and Yu.V. Nikitenko, Nuclear Instruments and Methods A 629, 245-250 (2011)
18. Yu.N. Khaydukov, V.L. Aksenov, Yu.V. Nikitenko et al, J. Supercond. and Novel Magn. 24, 961 (2011)
19. A. M. Nikitin, M.M. Borisov, E. Kh. Mukhamedzhanov et al, Cryst. Rep. 56, 858-865 (2011)
20. M. Tinkham, "Introduction to Superconductivity", McGraw-Hill, 1975
21. C.R. Hu, Phys. Rev. B 6, 1756 (1972)